\documentclass{aa}
\usepackage{epsf}
\usepackage{rotate}
\usepackage{graphics}
\usepackage{pslatex}

\begin{document}

\title{
      Kinematics, turbulence and evolution of planetary nebulae
      \thanks{Based on observations taken at the ESO }
      }

\author{
Krzysztof Gesicki,\inst{1} \and Agnes Acker,\inst{2} \and
Albert A. Zijlstra\inst{3} } 

\institute{
  Centrum Astronomii UMK, 
           ul.\,Gagarina 11, PL-87-100\, Torun, Poland \\
           \email{Krzysztof.Gesicki@astri.uni.torun.pl}
  \and
   Observatoire astronomique de Strasbourg, 11 rue de l'Universit\'e,
          67000 Strasbourg,  France\\
	  \email{acker@astro.u-strasbg.fr}
  \and
      Astrophysics Group, Physics Department, UMIST,
           PO Box 88, Manchester M60 1QD, United Kingdom \\
           \email{a.zijlstra@umist.ac.uk}
         }

\date{Received 23 October 2002 / Accepted 6 January 2003}

\abstract{

This paper discusses the location of a sample of planetary nebulae on
the HR diagram. We determine the internal velocity fields of 14
planetary nebulae from high-resolution echelle spectroscopy, with the
help of photoionization models. The mass averaged velocity is shown to
be a robust, simple parameter describing the outflow. The 
expansion velocity and radius are used to define the dynamical age;
together with the stellar temperature, this gives a measurement of the
luminosity and core mass of the central star. The same technique is
applied to other planetary nebulae with previously measured 
expansion velocities, giving a total sample of 73 objects. The objects
cluster closely around the Sch\"onberner track of 0.61\,M$_\odot$, with
a very narrow distribution of core masses. The masses are higher than
found for local white dwarfs. The luminosities determined in this way
tend to be higher by a factor of a few than those derived from the
nebular luminosities. The discrepancy is highest for the hottest (most
evolved) stars. We suggest photon leakage as the likely cause. The
innermost regions of the non-[WC] nebulae tend to show strong
acceleration. Together with the acceleration at the ionization front,
the velocity field becomes 'U'-shaped.  The presence of strong
turbulent motions in [WC] nebulae is confirmed. Except for this, we
find that the [WC] stars evolve on the same tracks as non-[WC] stars.

}

\maketitle

\section{Introduction}

The study of the expansion velocities of planetary nebulae (PN)
has two goals. The first goal is to develop our understanding of the
hydrodynamical evolution of the nebula. The importance of this is shown
by Stasi\'nska \&\ Tylenda (1994), who compare theoretical evolutionary
models with the planetary-nebula population of the Galactic Bulge. They
show how the comparison of such models with observed properties of the
PN population can be used to derive properties such as the mass
distribution of the central stars. However, they comment on the lack of
expansion velocity data, which is required to ``strongly constrain the
average properties of Galactic bulge PN and especially the masses of
their central stars''. 

The second goal is to characterize the parameters of the superwind and
study its dependence on the progenitor's age and metallicity. It is
expected that a dust-driven wind is less efficient at low metallicity,
leading to heavier white dwarfs (e.g. Zijlstra 1999), but there is no
observational confirmation of this. In fact white dwarf masses measured
in (metal-poor) globular clusters (Richer et al. 1997) seem low
compared to those of central stars of Bulge planetary nebulae (Gesicki
\&\ Zijlstra 2000).

In the present paper we discuss a sample of PNe for which we have
high-resolution spectra covering [\ion{O}{iii}], [\ion{N}{ii}] and
H$\alpha$, which allow us to model the velocity fields in some detail.
In two cases is the 6560\AA\ \ion{He}{ii} line used instead of
[\ion{O}{iii}].  Using these lines which trace both the inner and
outer regions, we find that the most common velocity profile is
'U'-shaped, with the highest velocities near the outer edge (as
predicted by hydrodynamical models) and the inner edge (which is less
expected).  The mass-averaged expansion velocity is shown to be a
well-defined and measurable parameter.  This parameter allows us to
accurately place the objects on the HR diagram.

\section{The observations}

The observations were performed at the ESO CAT telescope during 1993
and 1994.  This was a subsidiary 1.4m telescope feeding into the CES
spectrograph located at the neighbouring 3.6m telescope.  The long
camera was used giving a spectral resolution of 60\,000 (corresponding
to 5\,km\,s$^{-1}$).  The slit width was 2\arcsec, and the CCD was
binned to 2\arcsec pixels.  The spectra used here use only the central
row of pixels.  (As the nebulae studied here are compact and the CAT was
not an imaging-quality telescope, no spatial information was expected.)
The spectrum covers one order of the echelle, including H$\alpha$ and
the [\ion{N}{ii}] lines at 6548\AA\ and 6853\AA.  This was combined with
earlier published spectra of the [\ion{O}{iii}] 5007\AA\ line, take with
the same telescope but with a resolution of only 30\,000.

We assume the nebulae to be spherically symmetric. This is partly
necessitated by the data. The objects tend to be resolved by the
instrumental setup (as shown by the line splitting which is often
seen). But the CAT suffers from field rotation so that the orientation
of the slit on the sky changes during the exposure, and may be
different between the two separate setups. If the nebulae are
not symmetric, the two spectral settings may trace regions in
the nebula with different conditions.

For our modeling we will strictly assume spherical symmetry. {\it
Assuming} symmetry does not {\it make} the nebula symmetric---the most
common shape for PNe is in fact elliptical. In a previous paper
(Gesicki \&\ Zijlstra 2003) we discuss the effect of ellipticity on
the data and models, using very compact (extra-galactic) PNe for which
integrated spectra could be obtained. We will comment on this problem
below, when discussing the interpretation of the deduced velocity
profiles.

\section{Method of analysis}

Recently in Gesicki \&\ Zijlstra (2003) we published a very detailed
description of the modelling procedure, therefore we provide only a
short summary of the method.

The computer model first solves for the photo-ionization equilibrium,
where we attempt to reproduce the observed line intensities. Once
a satisfactory solution has been obtained, are the line profiles
calculated and compared with the observations.

The nebula is approximated as a spherical shell described by the inner
and outer radius, the total mass, the radial density distribution, the
radial velocity field, and the chemical composition. As far as possible
the values for the radii and chemical composition are taken from the
literature. The star is assumed to be a black body with  a  luminosity
and effective temperature. The temperature is again taken from the
literature if possible, while the luminosity of the star is determined
mainly by the H$\beta$ flux, once a distance has been chosen.

Once the photoionization model is complete, we calculate the
emissivity distribution for the ionic lines. The line profile can now
be calculated, taking into account the slit aperture (which does not
cover the full nebula) and the seeing (which scatters light rays
outside the aperture into the slit).  The velocity field is adjusted
to fit the line profiles of all three lines simultaneously. \ion{O}{iii}\
and \ion{N}{ii}\ tend to cover the inner and outer regions of the nebula,
respectively, while hydrogen is formed throughout the ionized nebula.
This set is therefore the minimum required for a realistic description
of the full nebula.

\section{The nebulae}

Spectra of 14 nebulae are analyzed in this paper. Twelve are located
towards the Galactic Bulge (but only 7 are likely Bulge members), and
two are in the Southern Galactic plane. The objects are specifically
chosen as having [\ion{O}{iii}] observations available.  More objects were
observed at the CAT in H$\alpha$ and [\ion{N}{ii}], but without [\ion{O}{iii}]
the velocity field can be insufficiently constrained.  There are
however two objects where a \ion{He}{ii} line was detected next to
H$\alpha$, and these are included in the present paper as this line
provides a good alternative to [\ion{O}{iii}].

We have previously analyzed a large set of nebulae using only the
[\ion{O}{iii}] line (Gesicki \&\ Zijlstra 2000). By necessity a simple,
linear velocity field was used. The addition of other lines allows us
to significantly improve on the earlier results.

The observed nebulae are listed in Table \ref{lista}, together with 
some parameters obtained from the fit. The spectra, together with
the fitted line profiles, are shown in Figs. \ref{0026}--\ref{3599}.

\subsection{Comments on individual nebulae}

\begin{figure*}
\resizebox{10cm}{!}{\includegraphics{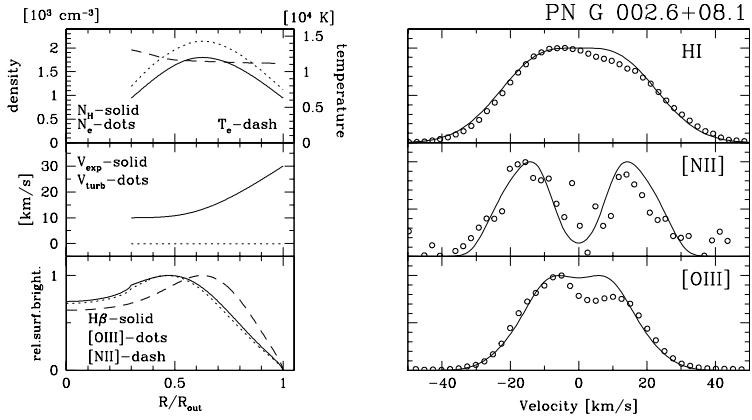}}
\caption{ The model results for the PN 002.6+08.1. The right
panels show the fit (solid lines) against the observations (open
circles) for all three lines. The top-left panel shows the density
distribution ($N_{\rm H}$ and $N_{\rm e}$) and the electron temperature.
The middle-left panel shows the velocity field and the bottom-left
panel shows the predicted surface brightness distribution (not including
seeing) which could be compared against radial cuts through images.
}
\label{0026}
\end{figure*}
\begin{figure*}
\resizebox{10cm}{!}{\includegraphics{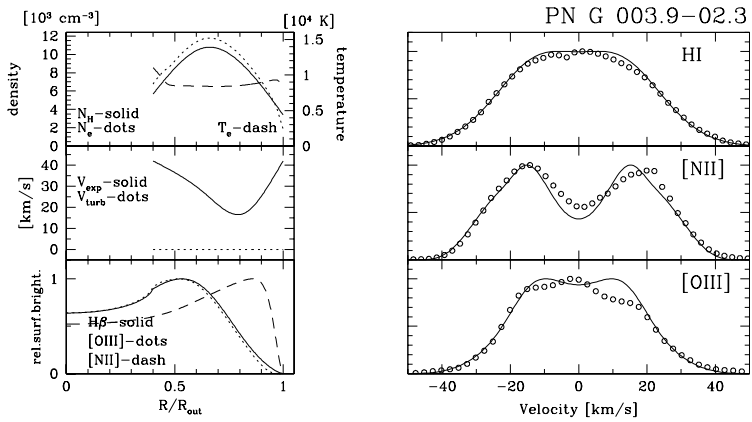}}
\caption{As Fig.\,\ref{0026}, for the PN 003.9-02.3.}
\label{0039}
\end{figure*}
\begin{figure*}
\resizebox{10cm}{!}{\includegraphics{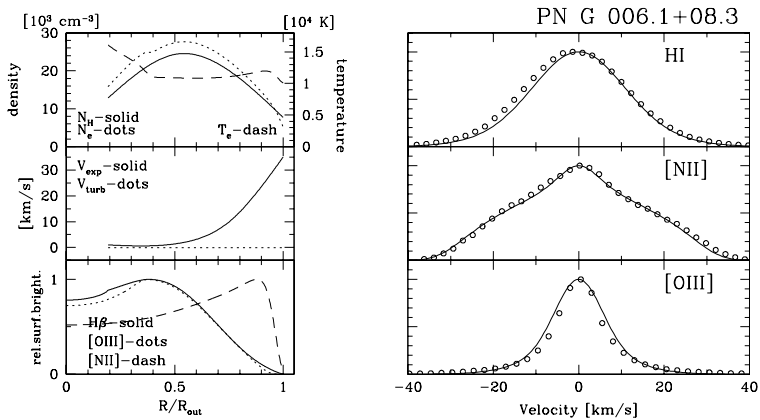}}
\caption{ As Fig.\,\ref{0026}, for the PN 006.1+08.3}
\label{0061}
\end{figure*}
\begin{figure*}
\resizebox{10cm}{!}{\includegraphics{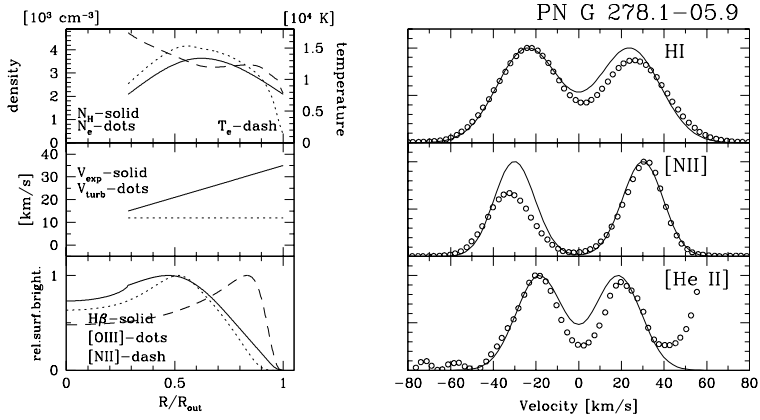}}
\caption{ As Fig.\,\ref{0026}, for the PN 278.1-05.9}
\label{2781}
\end{figure*}
\begin{figure*}
\resizebox{10cm}{!}{\includegraphics{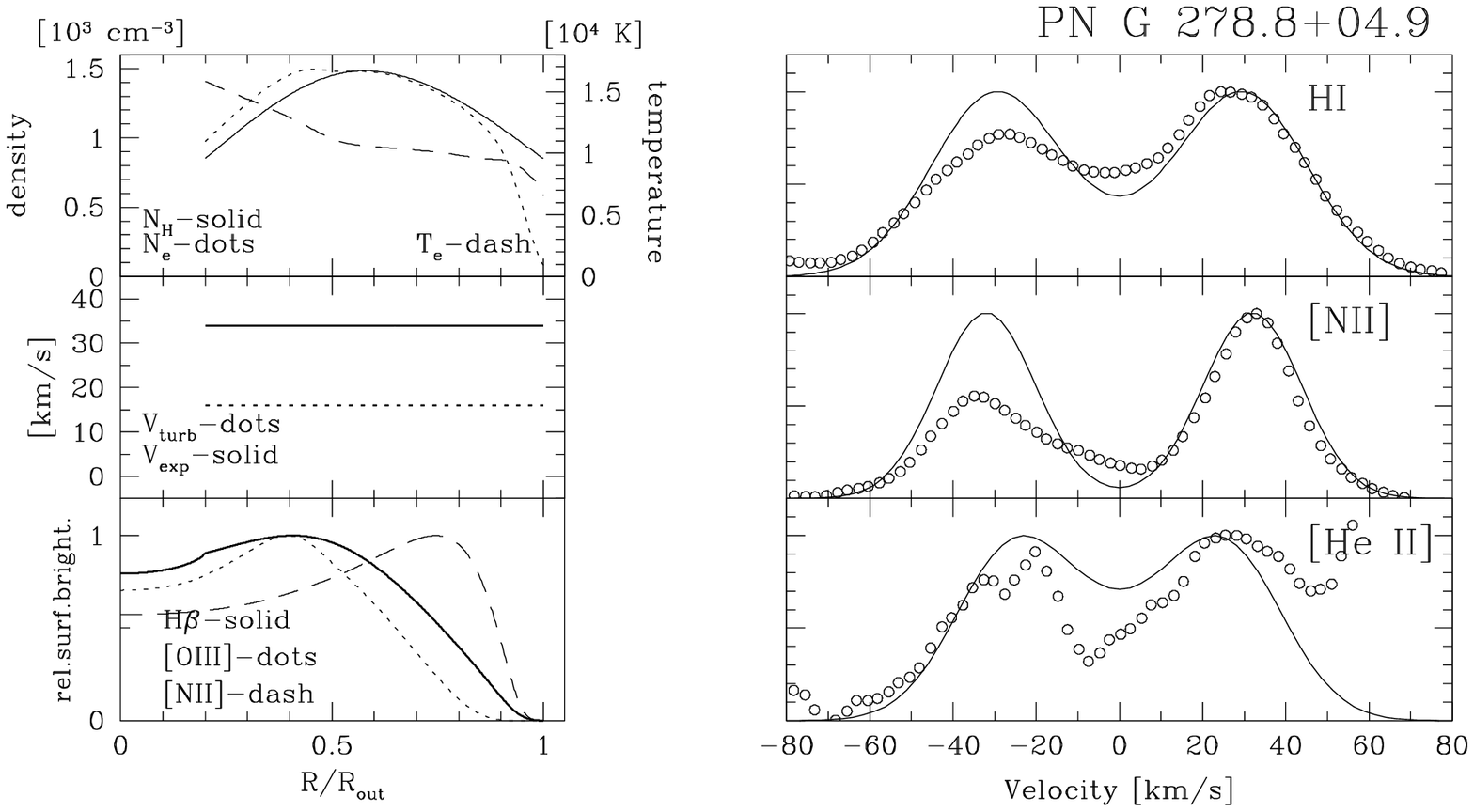}}
\caption{ As Fig.\,\ref{0026}, for the PN 278.8+04.9}
\label{2788}
\end{figure*}
\begin{figure*}
\resizebox{10cm}{!}{\includegraphics{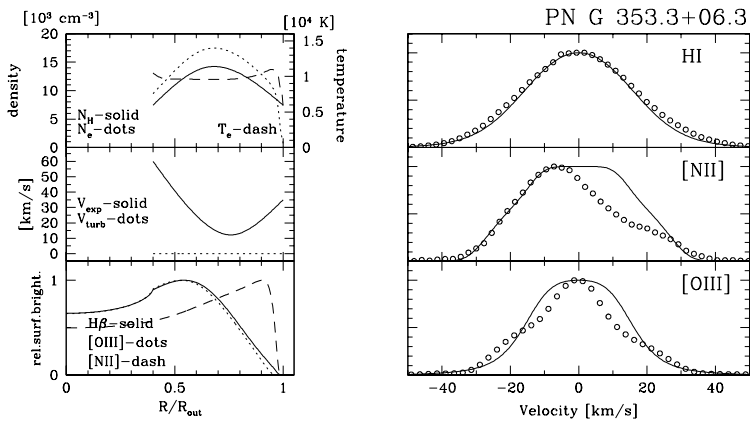}}
\caption{ As Fig.\,\ref{0026}, for the PN 353.3+06.3.}
\label{3533}
\end{figure*}
\begin{figure*}
\resizebox{10cm}{!}{\includegraphics{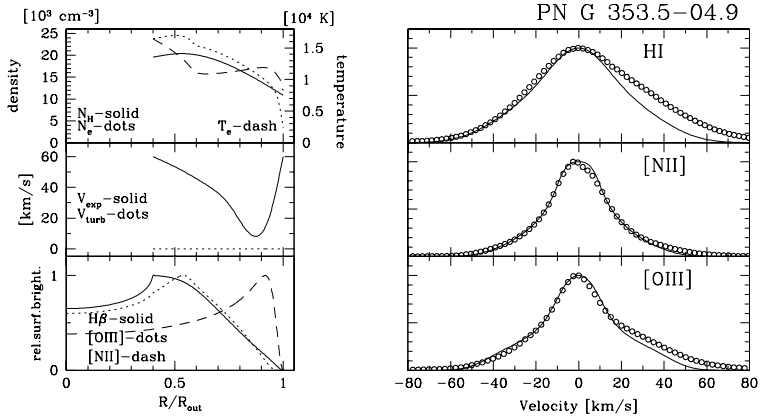}}
\caption{ As Fig.\,\ref{0026}, for the PN 353.5-04.9.}
\label{3535}
\end{figure*}
\begin{figure*}
\resizebox{10cm}{!}{\includegraphics{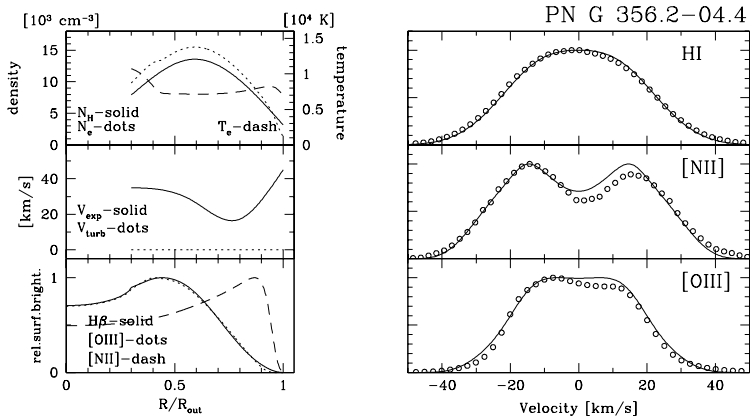}}
\caption{ As Fig.\,\ref{0026}, for the PN 356.2-04.4}
\label{3562}
\end{figure*}
\begin{figure*}
\resizebox{10cm}{!}{\includegraphics{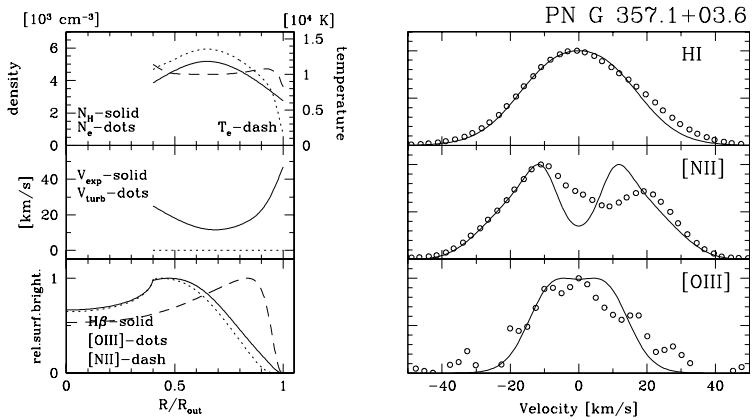}}
\caption{ As Fig.\,\ref{0026}, for the PN 357.1+03.6}
\label{3571}
\end{figure*}
\begin{figure*}
\resizebox{10cm}{!}{\includegraphics{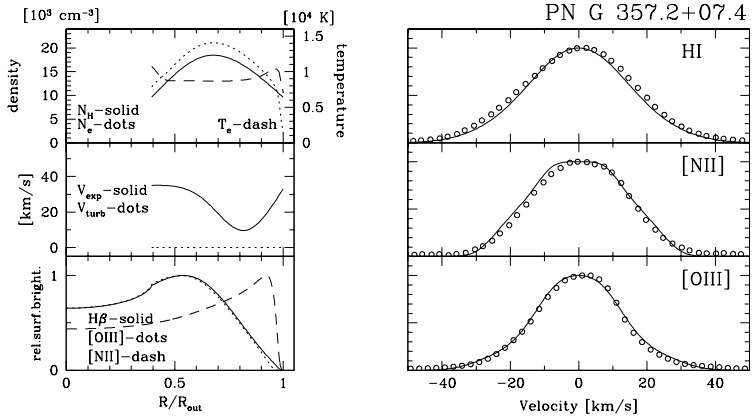}}
\caption{ As Fig.\,\ref{0026}, for the PN 357.2+07.4}
\label{3572}
\end{figure*}
\begin{figure*}
\resizebox{10cm}{!}{\includegraphics{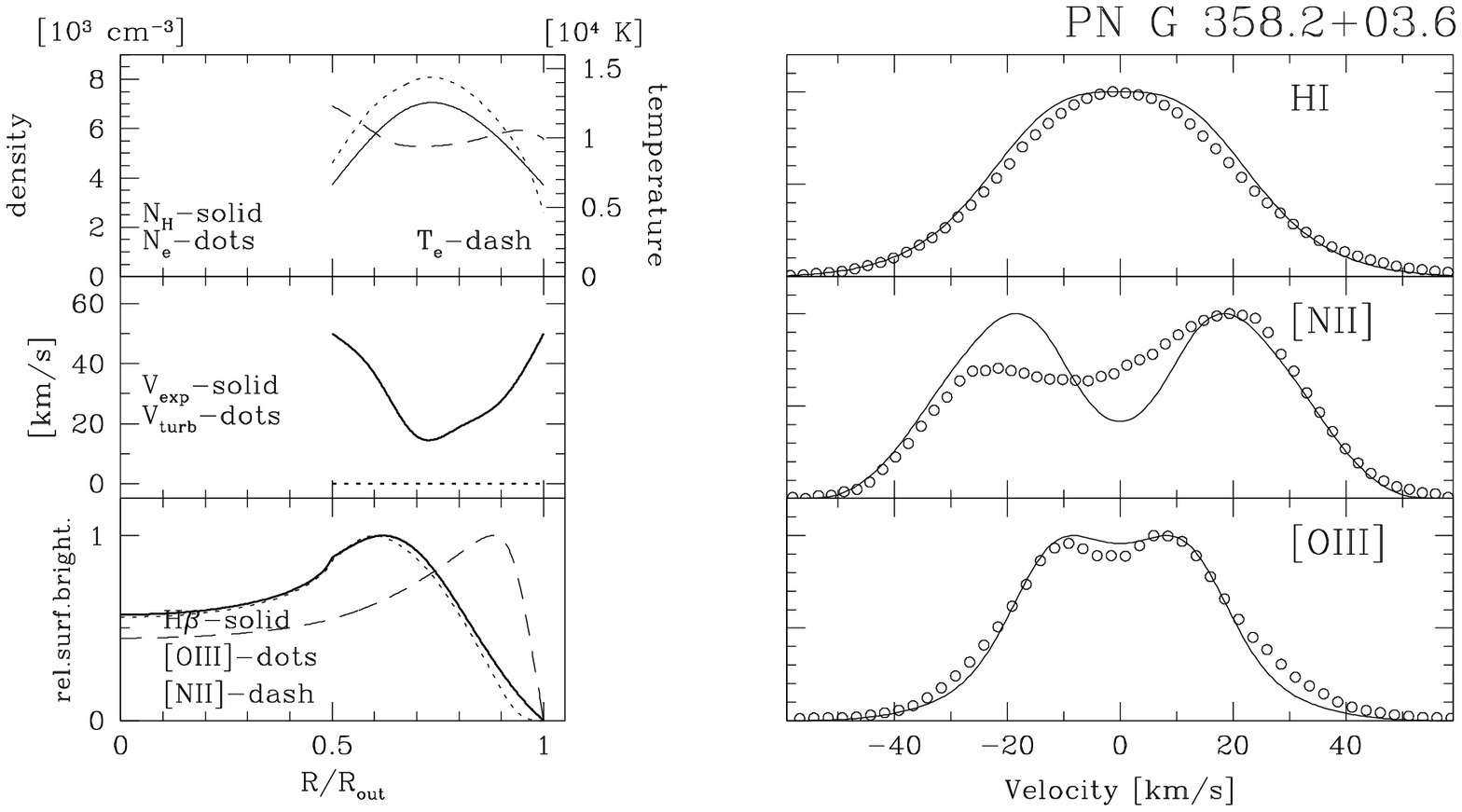}}
\caption{ As Fig.\,\ref{0026}, for the PN 358.2+03.6}
\label{3582}
\end{figure*}
\begin{figure*}
\resizebox{10cm}{!}{\includegraphics{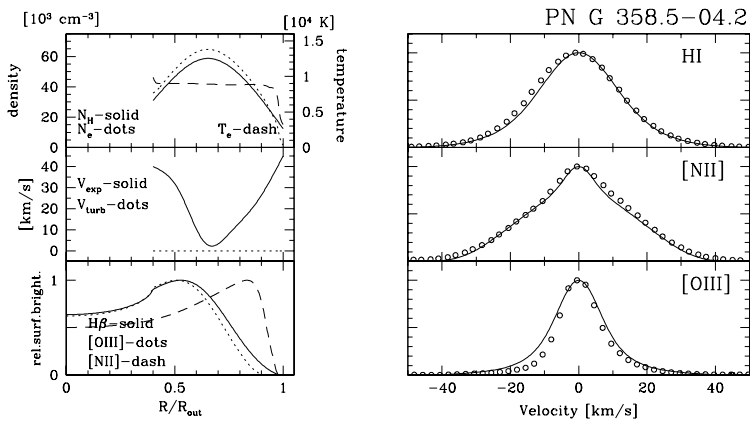}}
\caption{ As Fig.\,\ref{0026}, for the PN 358.5-04.2}
\label{3585}
\end{figure*}
\begin{figure*}
\resizebox{10cm}{!}{\includegraphics{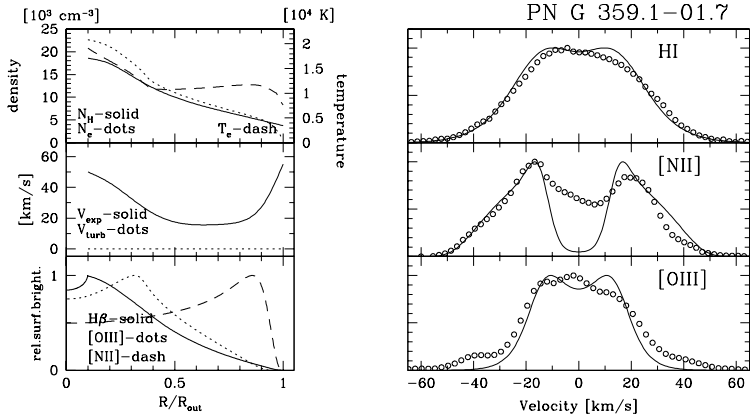}}
\caption{ As Fig.\,\ref{0026}, for the PN 359.1-01.7}
\label{3591}
\end{figure*}
\begin{figure*}
\resizebox{10cm}{!}{\includegraphics{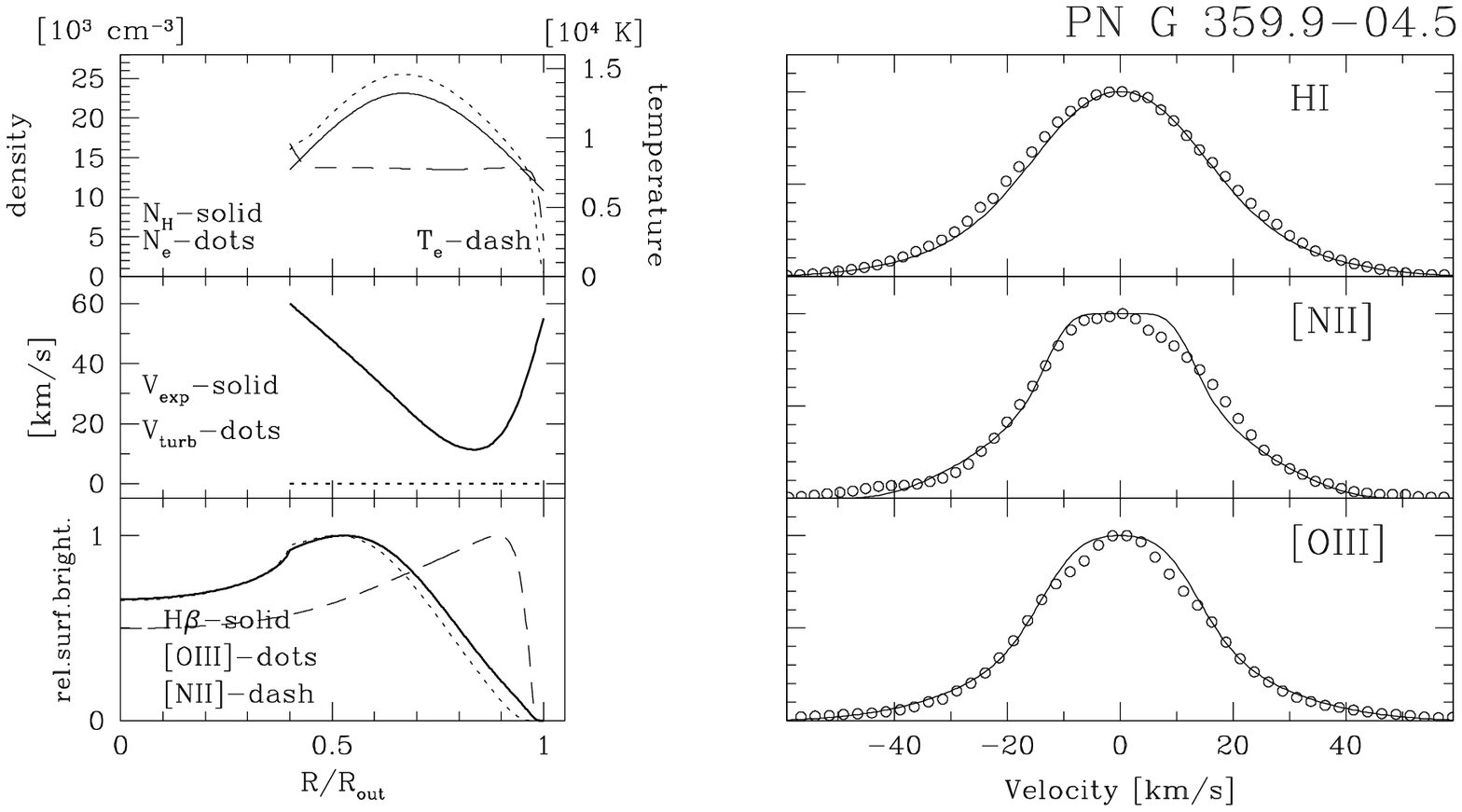}}
\caption{ As Fig.\,\ref{0026}, for the PN 359.9-04.5}
\label{3599}
\end{figure*}

{\it 002.6+08.1 (H 1-1) } This object, likely in the Bulge, is the only
density bounded object in our sample:  the evidence for this comes from
the weakness of the [\ion{N}{ii}] (and [\ion{S}{ii}]) lines.  The
electron temperature is high.  The inner radius is chosen as 0.3 (as
opposed to the usual 0.4) times the outer radius, in order to fit the
limited line splitting observed in [\ion{O}{iii}].  The fitted velocity
field increases monotonically with radius, with the velocity minimum at
the inner radius.  Compared to the other nebulae analyzed here, this
appears to be unusual.  There is no evidence for a turbulent component
in the velocity field, in spite of the wels-type (weak-emission-line
star).  The jump in the [\ion{N}{ii}] line profile at
$-20$\,km\,s$^{-1}$ (Fig.\,\ref{0026}) is believed to be an artifact of
the data reduction process.

{\it 003.9$-$02.3 (M 1-35) } The distance adopted for the fit is
inconsistent with Bulge membership: at larger distances the ionized
mass becomes rather high (0.36\,M$_{\odot}$).  A radio image is shown in
Zijlstra et al. (1989), but the resolution is limited. There is some
indication that the central peak is slightly shifted. The line
profiles show some structure. There is a dip near the centre of
H$\alpha$ which may be an instrumental problem. The [\ion{O}{iii}]
line show a central peak which is absent in the [\ion{N}{ii}]
line. This feature is not reproduced in the fit, but indicates excess
high-excitation gas at or near systemic velocity.  The best fit is
shown in Fig.\ref{0039}, using a 'U'-shaped velocity field, where the
minimum velocity is located at an intermediate radius (possibly near
the half-mass radius).

{\it 006.1+08.3 (M 1-20) } Schwarz et al.  (1992) have published an
image but it is difficult to measure a diameter from this.  The
high-resolution VLA radio image (Aaquist \&\ Kwok 1990) shows a
compact torus with very small inner radius (0.25$^{\prime\prime}$) and
outer radius of roughly 0.8$^{\prime\prime}$.  In the perpendicular
direction the outer radius is a little over 1$^{\prime\prime}$.  The
image appears mildly bipolar.  The fit uses an inner radius of $0.2
\times R_{\rm out}$, as indicated by the radio image.  The
[\ion{O}{iii}] line is very narrow.  The best fit therefore proposes
almost stationary gas (1\,km\,s$^{-1}$) at the inner radius.  The 
expansion velocity increases rapidly further out.

{\it 278.1$-$05.9 (NGC 2867)} This object contains a hot [WC] central
star.  
We have no [\ion{O}{iii}] data but the \ion{He}{ii} 6560\AA\ line is
detected and provides an alternative.  Schwarz et al.  (1992) present an
image showing a dense inner ring and a very large but much fainter halo.
Our fit, and probably our data, only relate to the inner ring.  The
strong [\ion{N}{ii}] and [\ion{S}{ii}] suggest that the inner ring is
ionization bounded.  This raises the question what ionizes the halo.
The optical image suggests the ring is clumpy:  photon leakage between
the clumps could ionize the halo---or the halo may be recombining.  The
fit suggests considerable turbulence, as appears to be the rule for [WC]
stars:  e.g.  Gesicki \&\ Zijlstra (2003) for a discussion of a similar
nebula.  In addition to the constant turbulence, the expansion velocity
increases outward.

{\it 278.8+04.9 (PB 6) } The photo-ionization model works poorly, as
the observed \ion{He}{ii} 4686\AA\ (Pe\~na et al. 2001) is far stronger
than the model predicts. However, the observed line may arise from the
star which is a hot [WC] star. As in the previous object, the \ion{He}{ii}
6561\AA\ was used instead of the [\ion{O}{iii}] line. A turbulent solution
provides the best fit, with a constant but high expansion velocity.

{\it 353.3+06.3 (M 2-6) } The [\ion{O}{iii}] line is very difficult to
fit, with a narrow core but wide and strong wings. The [\ion{N}{ii}] is
very asymmetric.  We have assumed that this reflects line splitting
where one component is very weak, but there may other interpretations.
In fact the radio image (Gathier et al. 1983) is almost unresolved, as
is the optical H$\alpha$ image of Bedding \&\ Zijlstra (1994),
suggesting line splitting is less likely.  The fit must be considered
as uncertain. However, the extended wings of the [\ion{O}{iii}] line
suggests high-velocity gas at or near the inner radius.

{\it 353.5$-$04.9 (H 1-36) } The [\ion{O}{iii}] has very wide
wings. The star is listed as a possible symbiotic (Acker et al. 1992).
Aaquist \&\ Kwok (1990) find the radio image to be unresolved, even
with the VLA (high resolution) A-array. The line profiles of both
H$\alpha$ and [\ion{O}{iii}] are asymmetric, with an extended red
wing. The compactness of the nebula indicates that this asymmetry is
intrinsic to the nebula. A 'U'-shaped velocity profile is used, with
velocity minimum close to the outer radius.  The outermost density is
also reduced to fit the rather narrow [\ion{N}{ii}] line.

{\it 356.2$-$04.4 (Cn 2-1) } The H$\alpha$ image of this object is
very compact (Bedding \&\ Zijlstra 1994). But our spectra show
significant splitting of the [\ion{N}{ii}] line, suggesting that the low
excitation gas is more extended than the slit width of
2$^{\prime\prime}$. The inner radius was chosen as $0.3 \times R_{\rm
out}$, to avoid line splitting of [\ion{O}{iii}].  A good fit could be
obtained with a 'U'-shaped velocity field.

{\it 357.1+03.6 (M 3-7) } The [\ion{O}{iii}] line is weak and rather noisy
which gives relatively poor constraints on the fit. The velocity of
the inner region is therefore not well constrained. The [\ion{N}{ii}] line
shows strong splitting whilst [\ion{O}{iii}] does not, but otherwise the
line widths are similar. The [\ion{N}{ii}] profiles require a component at
low velocity, but the [\ion{O}{iii}] is too wide for this component to
solely dominate the inner nebula. We therefore arrive at higher
velocities near the inner and the outer edge and lower in-between.
The poorly resolved image in Zijlstra et al. (1989) shows some
indication for ellipticity.

{\it 357.2+07.4 (M 4-3) } The lack of line splitting suggests an
unresolved nebula, in agreement with the radio image (Gathier et al.
1983) and H$\alpha$ image (Bedding \&\ Zijlstra 1994). The best fit
uses a 'U'-shaped velocity field with the highest velocity at the
inner edge.  The H$\alpha$ line is a little too narrow in the adopted 
fit, which may indicate that the electron temperature is underestimated.

{\it 358.2+03.6 (M 3-10) } A 'U'-shaped velocity field is required
with very high velocities (60\,km\,s$^{-1}$) near the inner and outer
edge.  The high velocity gas is indicated by the wide wings seen in all
three lines.  The [\ion{O}{iii}] line shows some splitting:  to
reproduce this, the inner radius was chosen as $0.5 \times R_{\rm out}$.

{\it 358.5$-$04.2 (H 1-46) } The radio image shows this to be a very
compact PN (Acker et al. 1992). (The optical image appears to suffer
from confusion with a nearby star: Bedding \&\ Zijlstra 1994).  A
'U'-shaped velocity field was used. However, the narrow [\ion{O}{iii}]
line is not well fitted.  The H-line is also a little too narrow in
the fit. Attempts to broaden this with a more complicated velocity
field, without affecting the other lines, did not work.

{\it 359.1$-$01.7 (M 1-29) } This is the most complicated object in the
sample.  The [\ion{O}{iii}] line shows several components including wings at
extreme velocities and an unresolved component at the systemic velocity.
The [\ion{N}{ii}] line show far less splitting in the observed profile than
in the fit.  These facts suggests that a bipolar outflow exists in this
object.  The radio image confirms that this object is bipolar (Zijlstra
et al.  1989).  Thus, the extreme wings probably result from the polar
flow, where the [\ion{O}{iii}] extends further into the nebula.  The
component at the systemic velocity could result from a near-stationary
torus (but could also be due to the slit having been centered on the
brightest component of the bipolar image).  A very small inner radius of
$0.1 \times R_{\rm out}$ was used, with a density maximum at the inner
edge.  Such a structure is needed because of the lack of line splitting
in the H$\alpha$ line, while the nebula is quite extended
(7$^{\prime\prime}$).  None of the lines are well fitted.  A spherically
symmetric model is probably a poor approximation to this nebula, and we
will not consider this object further.  However, as the bipolarity was
evident from the spectra, it appears likely that the other PNe in our
sample do not show such complicated structures, even if good images are
lacking.

{\it 359.9$-$04.5  (M 2-27) } The image in Gathier et al. (1983) shows
it to be almost unresolved.  A 'U'-shaped velocity gives a reasonable
fit. There are some residuals in the line cores of [\ion{O}{iii}]
and [\ion{N}{ii}].

\subsection{Summary}

\begin{table*}
\caption[]{The expansion velocities and other data concerning the 
nebula and its central star. In column 8, the number in braces
refers to the mass-averaged velocity found previously from fitting a
single line only.}
\begin{flushleft}
\begin{tabular}{ l l l l l l l l l  }
\cline{1-9}
\noalign{\smallskip} 
PN\,G & name & $\log\,T_{\rm eff}$ & $\log {L}
$ & dist. & R$_{\rm out}$ & M$_{\rm ion}$ & $V_{\rm av}$ & remarks \\
& & [K] &[L$_\odot$] & [kpc] & [pc] & [M$_{\odot}$] & [km\,s$^{-1}$] & 
\\ \noalign{\smallskip}
\cline{1-9}
\noalign{\smallskip}
002.6+08.1  & H 1-11     & 4.81 & 3.4  & 7.0  & 0.1   & 0.21  & 19 (22)  & wels \\ 
003.9$-$02.3  & M 1-35   & 4.85 & 3.6  & 4.5  & 0.05  & 0.138 & 25 (26)  &      \\ 
006.1+08.3  & M 1-20     & 4.90 & 3.4  & 6.0  & 0.026 & 0.045 & 12  (2)  &      \\ 
278.1$-$05.9  & NGC 2867 & 5.08 & 3.1  & 2.0  & 0.07  & 0.15  & 28       & [WO2] \\ 
278.8+04.9  & PB 6       & 5.04 & 2.7  & 4.0  & 0.10  & 0.18  & 34       & [WO1] \\ 
353.3+06.3  & M 2-6      & 4.74 & 3.7  & 8.4  & 0.04  & 0.103 & 22 (22)  &      \\ 
353.5$-$04.9  & H 1-36   & 5.08 & 3.5  & 7.0  & 0.03  & 0.061 & 32 (19)  &      \\ 
356.2$-$04.4  & Cn 2-1   & 4.90 & 3.6  & 6.0  & 0.04  & 0.087 & 25 (23)  & wels \\ 
357.1+03.6  & M 3-7      & 4.85 & 3.0  & 4.0  & 0.05  & 0.074 & 19       & wels \\ 
357.2+07.4  & M 4-3      & 4.80 & 3.62 & 8.0  & 0.033 & 0.075 & 20 (18)  &      \\ 
358.2+03.6  & M 3-10     & 4.97 & 3.0  & 5.0  & 0.04  & 0.048 & 27 (25)  &      \\ 
358.5$-$04.2  & H 1-46   & 4.70 & 3.9  & 7.0  & 0.02  & 0.045 & 17  (5)  &      \\ 
359.1$-$01.7  & M 1-29   & 5.08 & 3.5  & 3.0  & 0.05  & 0.128 & 24       & bipolar  \\ 
359.9$-$04.5  & M 2-27   & 4.78 & 3.7  & 5.5  & 0.03  & 0.069 & 26 (26)  &      \\ 
\noalign{\smallskip}
\cline{1-9}        
\end{tabular}       
\end{flushleft}     
\label{lista}
\end{table*}        

Of the 14 objects studies, 'U'-shaped velocity fields were required for
10.  Of the remaining four, two show significant turbulence.  Both
nebulae contain [WC]-type central stars:  turbulence around such stars
appears to be common (Acker et al.  2002) although the origin of the
turbulence is not clear (Mellema 2001).  Three nebulae, including one of
the turbulent objects, show monotonically increasing velocities.  The
prevalence of velocity maxima near the inner edge may be surprising.  It
was first found from the analysis of three PNe for which many emission
line profiles were measured, in deep echelle spectra (Gesicki \&\
Zijlstra 2003).  The present result suggests inner acceleration may be
quite common.  In earlier papers we only looked at monotonic velocity
fields.  The larger residuals in those fits may also be due to the
presence of high-velocity gas near the inner edge.

The data show that the 14 PNe of our sample have on average high
$T_{\rm eff}$ values, although within the usual temperature
range. This is due to the high stellar temperature needed to obtain
[\ion{O}{iii}] emission: at $T_{\rm eff} < 4\,10^4\,\rm K$, this line
is weak and these objects were not observed at the CAT.  This
selection effect should be taken into account when comparing the
present 14 PNe with the larger sample, it is clearly visible in the
figures discussed below, e.g. Fig.\,\ref{hr}.

\section{Discussion}

\subsection{Defining the expansion velocity}

The earlier analysis using the [\ion{O}{iii}] line only assumed a
linearly increasing velocity field. This often failed to fit the wings
of the [\ion{O}{iii}] line, but a more complicated velocity field
would be too poorly constrained. The present models show that most of
the nebulae reveal a 'U'-shape velocity field. Whether this field is
more common among PNe is hard to prove. Because of the selection
effect described above, the U-shaped velocity field may be a
consequence of a stronger wind blowing from the hotter stars in our
sample.

We compute the mass-averaged expansion velocity $V_{\rm av}$ as
described in Gesicki et al. (1998).  This single value was introduced
to characterize as simple as possible the nebular velocity field.
These values are given in Table\,\ref{lista}.  In the same column in
parentheses we show the $V_{\rm av}$ values obtained in Gesicki \&
Zijlstra (2000) from the [\ion{O}{iii}] line only with the assumption of
a linear velocity field. It is found that the $V_{\rm av}$ do not differ
much, except in two cases where previously we fitted the narrow line
core only, neglecting the extended wings. We conclude that the mass
averaged expansion velocity $V_{\rm av}$ is a well-defined
parameter. Although its value is best derived using several spectral
lines, a single line can give a reasonably good approximation to
$V_{\rm av}$, if the line shapes are not affected by perturbed
velocity fields. It appears that the sharp increases of the expansion
velocity towards the inner and outer nebular radius affect the line
wings shapes but do not alter much the main flow. Very often the fit
to the [\ion{N}{ii}] line wings requires a density decrease
simultaneously with the increase of velocity; this reduced the impact
of this region on $V_{\rm av}$. The situation is less clear at the
inner nebular radius, mostly because it is not well probed by the
emission lines in our sample, but the smaller volume in this region
contains a relatively small fraction of the nebular mass.

Our new data rather weaken the tendency found in Gesicki \& Zijlstra (2000)
that there seems to exist a weak correlation between the nebular outer
radius and $V_{\rm av}$. This is because the two lowest $V_{\rm av}$
values found in their paper are now revised upwards.

In the appendix we present a compilation of all PN for which 
expansion velocities have been determined. We note, however, that in
most of these only monotonic velocity fields were considered.  As
argued above, especially the average expansion velocity (weighted by
mass) appears to be relatively robust against such model
simplifications. In the next subsections the large sample of 73 objects
will be discussed. The smaller sample of 14 will be indicated in the
figures.

\subsection{The effect of metallicity}

The thermal broadening of hydrogen amounts to roughly
15\,km\,s$^{-1}$, which is similar to the typical expansion
velocities. For this reason the hydrogen lines are not so sensitive to
the details of the velocity fields and in previous papers have been
relatively easy to fit. But here we find that the models which best
reproduce O and N tend to fail at H$\alpha$, making the line either
slightly too narrow or too broad.

This may suggest that our photoionization model produces incorrect
electron temperatures, as this is the parameter which affects
hydrogen but to which the other, heavier atoms are much less sensitive.

The electron temperature is determined partly by the stellar 
temperature but most strongly by the metallicity. The reason is
that most of the cooling of the nebula occurs from the forbidden
oxygen lines. A lower abundance of oxygen leads to reduced cooling 
efficiency which is compensated by a higher electron temperature. 
We ran some test models, and found this to be the strongest effect in
changing the width of the hydrogen line.

We did not attempt to perform our own chemical composition analysis.

\subsection{Ionized masses}

\begin{figure}
\resizebox{\hsize}{!}{\includegraphics{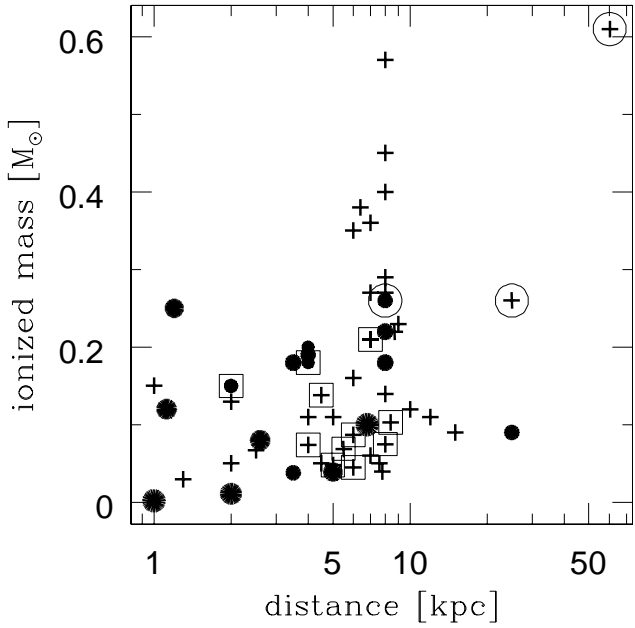}}
\caption{Ionized nebular mass versus distance for the PNe. All 
analysed earlier PNe are shown. The [WC]-type PNe are presented as circles
with their size proportional to their assigned [WC] class.
Other PNe are presented as plusses.
The objects analysed in the present article are framed with squares, and the three
objects
from Gesicki\& Zijlstra (2003) are framed with circles. The
distances are plotted in logarithmic scale to include the Sagittarius
and SMC nebulae}
\label{dm}
\end{figure}

One of the results of photoionization modelling is the amount of
ionized gas in the PN. This value is adopted to fit the observed and
dereddened $\log\,F({\rm H}\beta)$ at the assumed distance, size and
density distribution. It can be adjusted simultaneously with the
distance which is often poorly known. Sometimes it is possible to
build similarly satisfying models with different masses and distances:
the selection is made based on additional data. In Fig.\,\ref{dm} we
plot the derived ionized masses against the adopted distance. This
figure shows that most of our PNe have ionized masses below 0.2
M$_{\odot}$. The detailed analysis of Gesicki \&\ Zijlstra (2003) of
three PNe resulted in higher ionized masses than generally
expected. This finding supports our earlier results of adopting higher
masses for the Galactic Bulge nebulae. However it seems likely that
some of our PNe can be shifted between small distance \&\ low mass and
large distance \&\ high mass regions in the plot.

\subsection{Dynamical ages and core masses}

\begin{figure}
\resizebox{\hsize}{!}{\includegraphics{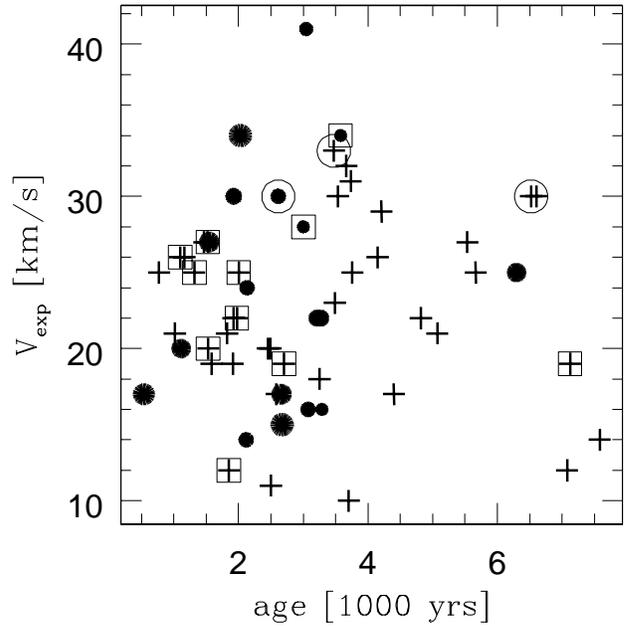}}
\caption{Nebular mass-averaged expansion velocity plotted against
dynamical age.  All PNe with
known metallicity are shown. The data are marked as in Fig.\,\ref{dm}.}
\label{va}
\end{figure}

\begin{figure}
\resizebox{\hsize}{!}{\includegraphics{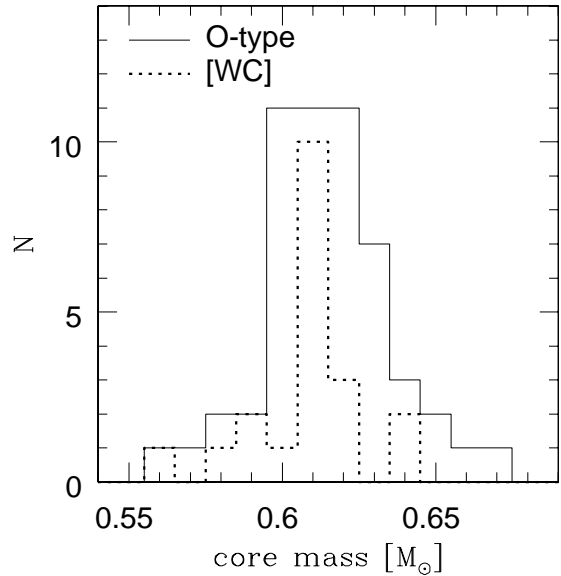}}
\caption{Histogram of the nebular core masses. All analysed earlier PNe
are included, the non-[WC] PNe (the solid line) and the [WC]-type PNe
(the dotted line) are plotted separately.}
\label{corm}
\end{figure}

The original AGB outflow velocity $V_{\rm AGB}$ can be estimated if
the luminosity and metallicity of the star is known.  Habing et
al. (1994) show that $V_{\rm AGB} \propto L_\ast^{0.33}\delta^{0.5}$
where $\delta$ is the dust-to-gas ratio, assumed to be proportional to
metallicity.  We have calculated (see Gesicki \&\ Zijlstra 2000)
$V_{\rm AGB}$ for the PNe with known metallicity (using the O/H in
Table\,\ref{listapp1}).  We assumed that a star with solar metallicity
(O/H=8.93) has an AGB outflow velocity of 15\,km\,s$^{-1}$. 

Gesicki \& Zijlstra (2000) describe a procedure to estimate the
dynamical ages of the PNe and masses of their central stars. We
follow their recipe for estimating the nebular dynamical age,
by calculating the average of the PN expansion velocity and the
original AGB velocity, and apply this value to a radius of 0.8 times
the outer radius. 

The mass-averaged expansion velocities we plot in Fig.\,\ref{va}
versus the dynamical age. We do not include objects for which the
metallicity is not known. There is no apparent correlation which
suggests that PNe do not speed up with time. Instead the increasing
ionization flux together with increasing stellar wind are used for
ionizing and accelerating more mass. The plot may suggest that
[WC]-type PNe are younger than average PNe.

To find the stellar parameters we applied the computer
codes to interpolate between the evolutionary tracks of Bl\"ocker
(1995) and Sch\"onberner (1983). These codes were written by
S.K.\,G\'orny and applied in G\'orny et al. (1997) for the derivation
of PN core masses. The basic idea of these programmes is to produce by
cubic spline interpolation a very dense grid of evolutionary
tracks. We adopted these routines to extract for each stellar
temperature, the luminosity and core mass against the evolutionary
age.  Then we interpolate the core mass and luminosity for each
object, based on its dynamical age and stellar temperature. Because
the dynamical age depends on the adopted AGB wind velocity, which
value is depending on stellar luminosity, we performed a simple
iteration.  At first we adopted the luminosities found from
photoionization modelling, then having located the objects on their
evolutionary tracks we recalculated the AGB velocities with new
luminosities and again interpolated stellar parameters.

The average core mass of the 14 PNe modelled in this paper is 0.62
M$_{\odot}$.  This is somewhat higher than the mean value obtained by
Gesicki \& Zijlstra (2000).  This difference can be explained again by
the previously mentioned selection effect, because more massive stars
evolve faster and can reach higher temperatures before the surrounding
nebula disperses.

For all PNe analyzed earlier we present the histograms of core masses
in  Fig.\,\ref{corm}. Both the average and the median values are now
0.61\,M$_{\odot}$. Separately we give the histogram for O-type
central stars and for those of [WC]-type.  Both histograms look
similar; however, the [WC]-type objects reveal a more pronounced peak at the
central position. This finding needs verification on a bigger sample.

Although the {\it relative} core masses are well determined, including
the narrow range in masses, the possibility of systematic effects
cannot be excluded. We assume that the acceleration of the nebula occurred
uniformly over time. If more time was spent at the original AGB velocity,
the dynamical age would be underestimated and the core mass overestimated.
We can quantify this by repeating our calculation using the AGB wind velocity
only, applied to the measured outer radius. This maximises the age and 
returns the lowest possible core mass. We find a systematic uncertainty
due to this effect of about $-0.015\,\rm M_\odot$. This gives a
{\it lower limit} to the mean core mass of about 0.6\,M$_\odot$.
The definition of zero position on evolutionary timescale also introduces
a systematic effect.

The evolutionary tracks are affected by the poorly known post-AGB
mass-loss rates. The tracks assume a mass loss tapering off after the
AGB, until replaced by a Pauldrach et al. (1988) wind. If the mass-loss rate
becomes similar to the nuclear burning rate ($\sim 10^{-7}\rm
\,M_\odot\,yr^{-1}$), the evolution in the HR diagram will accelerate
and the same time scales will be reproduced for lower core
masses. There is at present no evidence that mass-loss rates from
central stars of planetary nebulae reach such high values. For the
early post-AGB phase, the Bl\"ocker (1995) tracks use much higher
mass-loss rates than used by Vassiliadis \&\ Wood (1993); the latter
predicts much slower evolution which would give even higher core
masses.

The peak of the mass distribution of white dwarfs in the solar
neighbourhood is approximately 0.59\,M$_\odot$ (Napiwotzki et al.
1999). Although nominally discrepant, the difference can be due to
systematic effects, especially the effect of mass loss on the post-AGB
evolutionary tracks.  The mass loss in the early post-AGB phase may be
higher than assumed by Bl\"ocker (1995), or significant mass loss
event may happen during the PN evolution. The 'U'-shaped velocity
fields for which we find evidence could be indicative of a strong wind
during the PN phase, but it is not evident that this wind is massive
enough to affect the speed of the stellar evolution.

\subsection{Evidence for photon leakage}

\begin{figure}
\resizebox{\hsize}{!}{\includegraphics{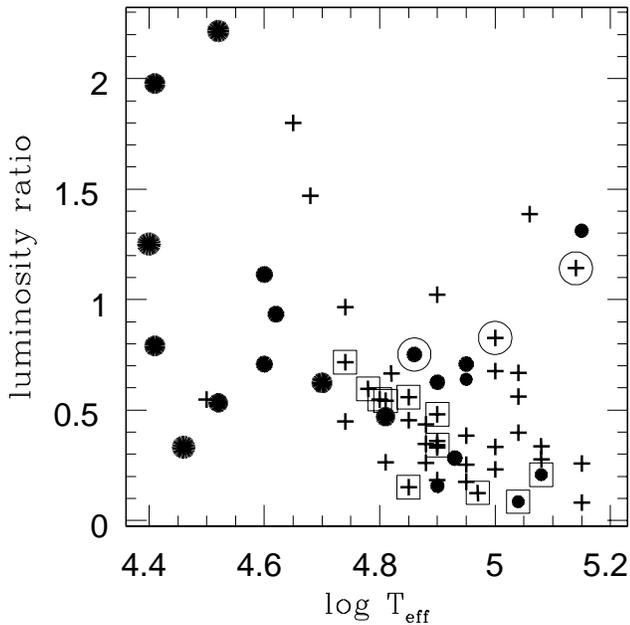}}
\caption{The ratio between the stellar luminosity obtained from 
photoionization modelling and the one obtained from evolutionary tracks
interpolation. The data are marked as in Fig.\,\ref{dm}.}
\label{lum}
\end{figure}

In Fig.\,\ref{lum} we plot for each object the ratio of the central
star luminosity obtained from photoionization modelling over the 
dynamical luminosity obtained from timescales and evolutionary tracks
interpolation. This ratio 
varies between 2 and 0.1, with the majority below 1. There is some
indication that the mean ratio declines with increasing temperature.
The values above unity probably reflect the uncertainty of the
ionization models, which for instance are quite sensitive to the distance.

Calculating the core masses for the limiting case of no post-AGB
acceleration, which gives lower luminosities, does not remove the
discrepancy between the luminosities. Taking larger distances can
bring the average values in agreement, but results in very large ionized
masses. Also, about half our sample is in the Bulge and has well-determined
(within 25\%) distances.

The stellar luminosity obtained from our modelling is the one needed
to ionize just the amount of mass needed to emit like the observed
nebula. The fact that the model luminosity is smaller than the value
predicted by evolutionary tracks may suggest that the nebula does not
intercept and transform all stellar ionizing flux. In other words, the
nebulae are leaking. This phenomenon seems to be stronger for the PNe
whose cores evolved to higher temperatures. In the sample of 14
nebulae, only a single object is found to be density bounded. In the
full sample, few nebulae are density bounded. The other objects
are ionization bounded and are not expected to 'leak' under the assumption
of spherical symmetry. The evidence for leaks points to asymmetries,
or clumping.

\subsection{The objects on the HR diagram}

\begin{figure*}
\resizebox{12cm}{!}{\includegraphics{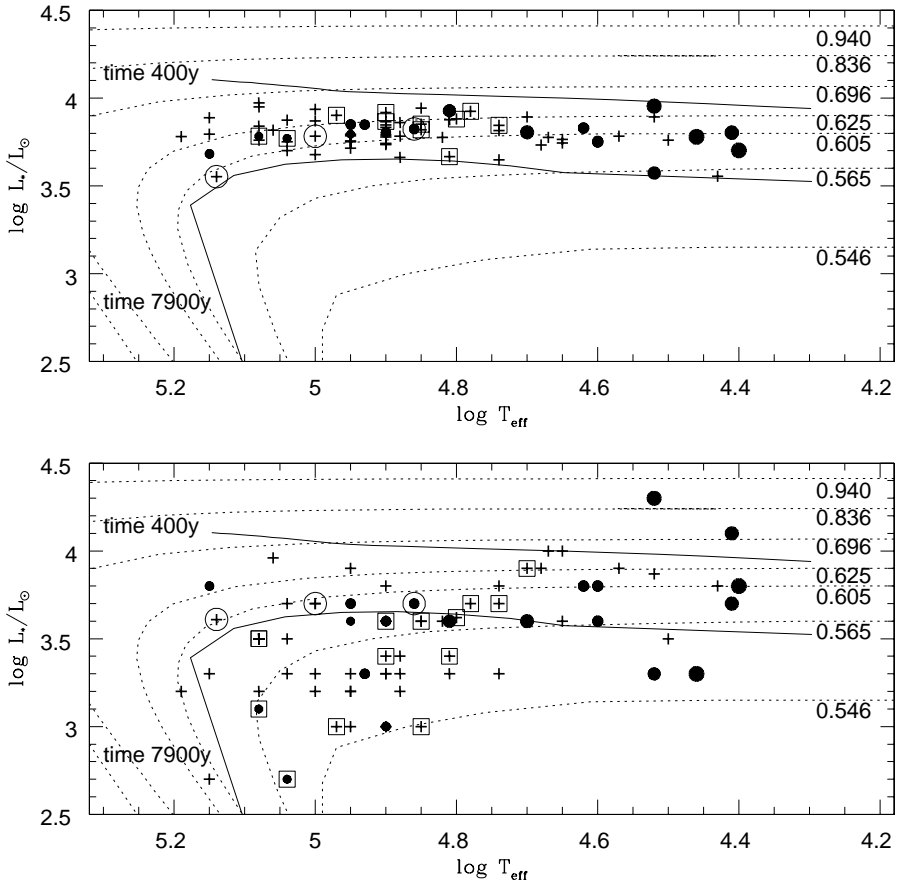}}
\caption{HR diagram for the PNe. All PNe analysed earlier are included.
The objects are marked as in Fig.\,\ref{dm}.
The upper panel corresponds to the dynamical luminosities, the lower to the
photoionization model luminosities (see text). The evolutionary tracks are 
overplot (dotted lines, labelled with core mass values), also are shown the 
lines corresponding to the smallest and the largest dynamical age of 
presented  sample}
\label{hr}
\end{figure*}

The procedure used in calculating the core masses also returns the
luminosity of the central star. Together with the central star
effective temperatures obtained from the photoionization models, we
can plot the PNe in a 'dynamical' HR diagram.  Fig.\,\ref{hr} combines
all objects from our earlier publications, the PNe analysed in this
paper (framed with squares), and the three PNe of Gesicki \&\ Zijlstra
(2003) (framed with circles). Because we have two estimations of
stellar luminosities we present the data in two boxes.

The objects show a strong clustering around a single evolutionary
track.  The fact that the objects agree fairly well with a single
track across a range of temperatures suggests that the relative speed
of evolution of the tracks is probably correct. If, for instance, the
evolution would speed up at a certain temperature, this would appear as
a tilted observed sequence compared to the theoretical tracks.  Such a
tilt is not evident. 

We have no objects on the cooling track. The knee in the HR diagram is
traversed very quickly (Bl\"ocker 1995) so that few objects would be
expected there.  Further down on the cooling track the PN become very
faint.  However, near the knee the procedure to determine the mass and
luminosity as a function of temperature and age has two solutions.  We
select the more luminous one and therefore may have missed one or more
objects on the cooling track.  The luminosity independently determined
in the ionization model provides a check:  in only one case (M2-11, a
Bulge PN) could this place the object low on the cooling track.

The lower panel of Fig.\,\ref{hr} shows the same diagram but now using
the luminosity from the photoionization model.  Comparing the two
panels shows that the model luminosities do not cluster in the
diagram, and are almost always lower than the dynamical
luminosities. The lower panel also shows a characteristic decline of
luminosity with increasing temperature, in contradiction to the
horizontal tracks. This decline has often been seen in nebular
luminosities, e.g. Pottasch (1984). Comparison with the dynamical
luminosities suggest that measurements of the nebulae (e.g., using the
relation $L_\ast = 150 L_{\rm H \beta}$, Gathier \& Pottasch 1989,
Zijlstra \&\ Pottasch 1989, Magrini et al. 2002), especially with
hotter stars, underestimate the stellar luminosity.

\subsection{[WC]-type planetary nebulae}

In all figures of section 5 we separated the [WC]-type PNe from
other types to check how this group of objects behaves.  These PNe are
characterised by the presence of a hydrogen-deficient central star,
which has a much stronger (10--100 times) stellar wind than other PNe
cores. Approximately 8\%\ of PNe have [WC]-type stars (G\'orny 2001).
The nebulae surrounding [WC]-type and O-type stars are, however,
very similar. The work of Gesicki \&\ Acker (1996) and Acker et
al. (2002) discusses a significant difference between the complete two
groups: [WC]-type PNe are characterised by strong turbulent motions
while the other are not. 

Our present work confirms the limited observable differences concerning
the turbulent motions.  The distribution of stellar temperatures is
different, with the [WC] stars being on average some 20\,000 K
cooler--they dominate our sample at the coolest temperatures. However,
they were taken from a different sample and this difference can be
explained by selection effects. Otherwise, the expansion
velocities, ionized masses, the histogram of derived central star
masses and the distribution in the HR diagram are the same between the
[WC] and non-[WC] stars. From the presented HR diagram it looks very
likely that the [WC]-type stars follow the same evolutionary tracks as
the other PNe, and are derived from the same progenitor population.

The fact that turbulence is an exclusive property of [WC] stars shows
that evolution from [WC] star to non-[WC] star must be rare at best,
and that the known [WC] stars spend a sufficient amount of time in this
phase for their winds to impart sufficient kinetic energy to the
nebulae to cause the observed turbulence. The nebular ages are also no
different between [WC] and non-[WC] stars.  All these factors suggest that
although the two groups arise from the same stellar population, the
[WC] stars separate relatively early in the post-AGB evolution. The
model where the strong wind strips the hydrogen layer of a helium
burning star can explain the observed correlation between [WC] stars
and turbulence, as well as the otherwise very similar nebulae. 

\section{Conclusions}

We have presented the expansion velocity fields in a sample of 14
planetary nebulae. The nebulae have high resolution spectra
covering for each object three emission lines probing different
nebular layers.

Two PNe show significant turbulence and both contain [WC]-type central
stars. This confirms the existing  evidence that
turbulence around such stars is common (Acker et al. 2002).

Three nebulae, including one of the turbulent objects, show 
monotonically increasing velocities. 

For ten objects (nonturbulent) we derived a U-shaped velocity field.
The outer velocity increase is a well-understood feature of the
ionization front. The inner velocity increase was postulated
recently by Gesicki \& Zijlstra (2003) for two PNe with  very hot
cores (above 100\,000\,K). We explained this in terms of a strong wind
blowing from a hot central star and sweeping up the inner nebular layers.
We thought of it as an exception  but the presented results suggests
that the inner acceleration may be quite common. It is already present
at $T_{\rm eff} \sim 50\,000\,{\rm K}$ which is the lowest value in our
sample.

We found that the mass-averaged expansion velocity is a reasonably
robust parameter. It is well determined even from a single emission
line, if this line doesn't exhibit unusual features.  The recently
derived U-shape velocity fields result in much improved fits to the
line wings, but the $V_{\rm av}$ are almost the same as obtained with
simple linearly increasing velocity.  We conclude that $V_{\rm av}$ is
the proper parameter to describe the main nebular flow.

In the discussion we assembled all previously analysed  nebulae. This
has not been done before. 24 objects have been discussed by Acker et
al. (2002); those PNe were observed in three or more emission lines. 44
objects have been discussed by Gesicki \& Zijlstra (2000), observed
only in the [\ion{O}{iii}] 5007\AA\ line. Because we find  that the
mass-averaged expansion velocity is a robust parameter, we decided to
merge the two samples and add three PNe from Gesicki \& Zijlstra
(2003).

Applying the interpolation routines for evolutionary tracks, we
estimate the nebular core masses. The values for our sample of 14 are
somewhat larger than the mean value for the full sample. The mean
value for the whole sample of 73 PNe is 0.61\,M$_{\odot}$ which is
higher than the expected average white dwarf mass. This difference can
possibly be explained in terms of systematic effects.

For all discussed PNe we also interpolated the predicted by
evolutionary calculation stellar luminosities. We find that for the
14 PNe and for the majority of the remaining nebulae, our stellar
luminosities are higher than the values obtained from the
photo-ionization model of the nebula.  We conclude that PNe do not
intercept all ionizing radiation even when there is strong evidence
of the presence of an ionization boundary.

All analysed PNe show a strong clustering around a single evolutionary
track on the HR diagram. This provides support for the evolutionary
calculations.

In the discussion we separated the [WC]-type PNe from other types. In
all plots presented the distributions of the two groups are actually
the same. It looks like the [WC]-type stars follow the same
evolutionary track as the other PNe cores. However, their true
position on the HR diagram may be masked by pseudo-photospheres in the
strong wind. 

\section*{Acknowledgements}

The evolutionary tracks interpolation routines were kindly provided by
S.K.G\'orny. This project was financially supported by the CNRS through
the JUMELAGE  ``Astronomie Pologne'' programme, by the ``Polish State
Committee for  Scientific Research'' through the grant No.
2.P03D.020.17 and by a PPARC visitor grant PPA/G/O/2001/00483.


\appendix

\section{Full sample of mass-averaged expansion velocities}

We present here the data concerning the full sample of PNe with
mass-averaged expansion velocities derived. The table contains in
consecutive columns: the PNG designation, the usual name,
stellar effective temperature,  stellar luminosity derived from
photoionization models, assumed distance, physical outer radius, model's
ionized mass, mass-averaged expansion velocity, additional turbulent
velocity, O/H abundance ratio, nebular dynamical age, interpolated
central star mass, remarks.

\begin{table*}
\caption[]{The expansion velocities and other data concerning the 
nebula and its central star. Part 1.}
\begin{flushleft}
\begin{tabular}{ l l l l l l l l l l l l l }
\cline{1-13}
\noalign{\smallskip} 
PN\,G & name & $\log\,T_{\rm eff}$ & $\log {L} $ & dist. & $R_{\rm out}$ & $M_{\rm ion}$ & $V_{\rm av}$   &
                   $V_{\rm turb}$ & O/H & $t_{dyn}$ & $M_*$            & remarks \\
      &      & [K]                 &[L$_\odot$]  & [kpc] & [pc]        & [M$_{\odot}$]   & [km\,s$^{-1}$] & 
                   [km\,s$^{-1}$] &     & [$10^3$y] &  [M$_{\odot}$]   &       \\ 
\noalign{\smallskip}
\cline{1-13}
\noalign{\smallskip}
001.5-06.7 & SwSt1      &  4.52  &  4.30  &  2.00  & 0.007  &  0.011  & 17.0    &  14  &  7.50   &    0.53   &   0.640 &  WC10  \\
001.7-04.6 & H1-56      &  4.95  &  3.00  &  10.0  &  0.08  &  0.12   & 17.0    &  0   &  8.73   &    4.40   &   0.606 & \\
002.0-06.2 & M2-33      &  4.74  &  3.30  &  8.00  &  0.10  &  0.27   & 12.0    &  0   &  8.69   &    7.07   &   0.574 & \\
002.0-13.4 & IC4776     &  4.60  &  3.60  &  3.50  &  0.06  &  0.18   & 22.0    &  0   &  8.25   &    3.27   &   0.594 &  WC6 \\
002.1-02.2 & M3-20      &  4.88  &  3.20  &  7.00  &  0.10  &  0.27   & 32.0    &  0   &  8.64   &    3.66   &   0.607 & \\
002.6+08.1 & H1-11      &  4.81  &  3.40  &  7.00  &  0.10  &  0.21   & 19.0    &  0   &  7.60   &    7.12   &   0.579 & \\
003.1+02.9 & Hb4        &  4.95  &  3.6   &  4.0   & 0.065  &  0.2    & 16.0    &  14  &  8.92   &    3.28   &   0.612 &  WC3 \\
003.2-06.2 & M2-36      &  4.90  &  3.30  &  6.40  &  0.11  &  0.38   & 22.0    &  0   &  8.89   &    4.81   &   0.601 & \\
003.7-04.6 & M2-30      &  5.04  &  3.50  &  7.00  &  0.10  &  0.36   & 26.0    &  0   &  8.74   &    4.15   &   0.610 & \\
003.8-17.1 & Hb8        &  4.88  &  3.40  &  15.0  &  0.04  &  0.09   & 22.0    &  0   &  8.58   &    1.92   &   0.622 & \\
003.9-02.3 & M1-35      &  4.85  &  3.60  &  4.50  &  0.05  &  0.138  & 25.0    &  0   &  8.83   &    2.00   &   0.619 & \\
004.0-03.0 & M2-29      &  4.88  &  3.30  &  9.00  &  0.08  &  0.23   & 14.0    &  0   &  7.46   &    7.58   &   0.581 & \\
004.2-04.3 & H1-60      &  4.90  &  3.00  &  7.00  &  0.10  &  0.21   & 21.0    &  0   &  8.61   &    5.07   &   0.599 &  WC? \\
004.8-22.7 & He2-436    &  4.95  &  3.70  &  25.0  &  0.03  &  0.09   & 14.0    &  0   &  8.36   &    2.11   &   0.622 &  WC5 \\
004.9+04.9 & M1-25      &  4.62  &  3.8   &  8.0   &  0.06  &  0.18   & 30.0    &  12  &  9.09   &    1.92   &   0.611 &  WC6 \\
005.2+05.6 & M3-12      &  5.0   &  3.2   &  8.0   &  0.15  &  0.57   & 30.0    &  0   &  8.14   &    6.60   &   0.595 & \\
005.8-06.1 & NGC6620    &  5.04  &  3.3   &  8.0   &  0.14  &  0.45   & 27.0    &  0   &  8.84   &    5.53   &   0.603 & \\
006.0-03.6 & M2-31      &  4.86  &  3.7   &  8.00  &  0.07  &  0.26   & 30.0    &  0   &  8.7    &    2.61   &   0.614 &  wels \\
006.1+08.3 & M1-20      &  4.90  &  3.40  &  6.00  & 0.026  &  0.045  & 12.0    &  0   &  8.53   &    1.85   &   0.624 & \\
006.8-19.8 & Wray16-423 &  5.00  &  3.70  &  25.0  &  0.09  &  0.26   & 33.0    &  0   &  8.33   &    3.47   &   0.613 &  wels \\
006.8+04.1 & M3-15      &  4.90  &  3.6   &  4.0   &  0.06  &  0.19   & 16.0    &  15  &  8.89   &    3.07   &   0.612 &  WC4-6 \\
008.3-01.1 & M1-40      &  5.15  &  3.3   &  2.5   & 0.042  &  0.067  & 27.0    &  0   &  9.02   &    1.46   &   0.648 & \\
009.4-09.8 & M3-32      &  4.95  &  3.20  &  7.00  &  0.10  &  0.26   & 23.0    &  0   &  0      &    4.53   &   0.605 & \\
009.6-10.6 & M3-33      &  4.95  &  3.30  &  8.00  &  0.12  &  0.40   & 25.0    &  0   &  8.45   &    5.66   &   0.598 & \\
010.7-06.4 & IC4732     &  4.82  &  3.60  &  8.00  &  0.08  &  0.29   & 25.0    &  0   &  8.42   &    3.75   &   0.604 & \\
027.6+04.2 & M2-43      &  4.81  &  3.6   &  5.0   &  0.02  &  0.039  & 20.0    &  10  &  8.30   &    1.11   &   0.636 &  WC8 \\
029.2-05.9 & NGC6751    &  4.90  &  3.0   &  2.0   &  0.1   &  0.15   & 41.0    &  15  &  8.60   &    3.04   &   0.612 &  WC4 \\
064.7+05.0 & BD+30      &  4.41  &  4.1   &  2.6   & 0.035  &  0.08   & 27.0    &  15  &  8.40   &    1.54   &   0.606 &  WC9 \\
096.4+29.9 & NGC6543    &  4.70  &  3.6   &  1.12  &  0.05  &  0.12   & 17.0    &  12  &  8.75   &    2.66   &   0.608 &  WC9 \\
120.0+09.8 & NGC40      &  4.52  &  3.3   &  1.2   &  0.14  &  0.25   & 25.0    &  8   &  8.70   &    6.29   &   0.564 &  WC8 \\
123.6+34.5 & IC3568     &  4.68  &  3.9   &  2.0   &  0.1   &  0.13   & 29.0    &  0   &  8.43   &    4.20   &   0.590 &  O3(H) \\
144.5+06.5 & NGC1501    &  5.15  &  3.8   &  1.2   &  0.16  &  0.18   & 40.0    &  10  &  0      &    4.17   &   0.614 &  WC4 \\
146.7+07.6 & M4-18      &  4.4   &  3.8   &  6.8   &  0.06  &  0.1    & 15.0    &  15  &  9.23   &    2.67   &   0.580 &  WC11 \\
215.2-24.2 & IC418      &  4.57  &  3.9   &  1.0   & 0.036  &  0.05   & 15.0    &  0   &  0      &    2.50   &   0.602 &  Of(H) \\
221.3-12.3 & IC2165     &  5.19  &  3.2   &  2.5   & 0.057  &  0.076  & 25.0    &  14  &  0      &    2.37   &   0.632 &  wels? \\
261.0+32.0 & NGC3242    &  4.9   &  3.3   &  1.0   &  0.1   &  0.15   & 31.0    &  0   &  8.66   &    3.73   &   0.608 &  O(H) \\ 
278.1-05.9 & NGC2867    &  5.08  &  3.1   &  2.0   &  0.07  &  0.15   & 28.0    &  12  &  8.44   &    3.00   &   0.619 &  [WO2] \\
278.8+04.9 & PB6        &  5.04  &  2.7   &  4.0   &  0.1   &  0.18   & 34.0    &  16  &  8.57   &    3.57   &   0.614 &  [WO1] \\
\noalign{\smallskip}
\cline{1-13}        
\end{tabular}       
\end{flushleft}     
\label{listapp1}
\end{table*}        

\begin{table*}
\caption[]{The expansion velocities and other data concerning the 
nebula and its central star. Part 2.}
\begin{flushleft}
\begin{tabular}{ l l l l l l l l l l l l l }
\cline{1-13}
\noalign{\smallskip} 
PN\,G & name & $\log\,T_{\rm eff}$ & $\log {L} $ & dist. & $R_{\rm out}$ & $M_{\rm ion}$ & $V_{\rm av}$   &
                   $V_{\rm turb}$ & O/H & $t_{dyn}$ & $M_*$            & remarks \\
      &      & [K]                 &[L$_\odot$]  & [kpc] & [pc]        & [M$_{\odot}$]   & [km\,s$^{-1}$] & 
                   [km\,s$^{-1}$] &     & [$10^3$y] &  [M$_{\odot}$]   &       \\ 
\noalign{\smallskip}
\cline{1-13}
\noalign{\smallskip}
285.4+01.5 & Pe1-1      &  4.93  &  3.3   &  3.5   &  0.05  &  0.038  & 24.0    &  10  &  8.75   &    2.13   &   0.621 &  WC4-5 \\
296.3-03.0 & He2-73     &  5.00  &  3.70  &  4.00  &  0.04  &  0.11   & 19.0    &  0   &  8.80   &    1.91   &   0.629 & \\
307.2-09.0 & He2-97     &  4.81  &  3.30  &  4.50  &  0.03  &  0.05   & 19.0    &  0   &  8.58   &    1.58   &   0.624 & \\
327.1-02.2 & He2-142    &  4.41  &  3.7   &  3.5   &  0.03  &  0.03   & 20.0    &  7   &  0      &    1.56   &   0.605 &  WC9 \\
327.5+13.3 & He2-118    &  4.90  &  3.00  &  7.00  &  0.03  &  0.03   & 13.0    &  0   &  0      &    2.40   &   0.617 & \\
345.0-04.9 & Cn1-3      &  4.65  &  3.60  &  7.50  &  0.05  &  0.12   & 16.0    &  0   &  0      &    3.25   &   0.598 & \\
345.2-08.8 & Tc1        &  4.5   &  3.5   &  2.0   &  0.05  &  0.05   & 20.0    &  0   &  8.71   &    2.49   &   0.596 &  Of(H) \\
346.3-06.8 & Fg2        &  5.08  &  3.20  &  8.70  &  0.10  &  0.22   & 30.0    &  0   &  8.91   &    3.53   &   0.615 & \\
347.4+05.8 & H1-2       &  5.06  &  3.96  &  7.00  &  0.03  &  0.10   & 11.0    &  0   &  8.33   &    2.49   &   0.623 & \\
349.8+04.4 & M2-4       &  4.74  &  3.80  &  6.00  &  0.05  &  0.16   & 17.0    &  0   &  8.80   &    2.58   &   0.610 & \\
350.9+04.4 & H2-1       &  4.52  &  3.87  &  5.00  &  0.03  &  0.06   & 36.0    &  0   &  0      &    0.86   &   0.624 & \\
351.9+09.0 & PC13       &  4.95  &  3.90  &  6.00  &  0.10  &  0.15   & 29.0    &  0   &  0      &    3.59   &   0.610 & \\
352.9-07.5 & Fg3        &  4.67  &  4.00  &  7.00  &  0.03  &  0.09   & 10.0    &  0   &  0      &    3.12   &   0.601 & \\
352.9+11.4 & K2-16      &  4.46  &  3.3   &  1.0   &  0.05  &  0.002  & 34.0    &  12  &  7.90   &    2.02   &   0.600 &  WC11 \\
353.3+06.3 & M2-6       &  4.74  &  3.70  &  8.40  &  0.04  &  0.103  & 22.0    &  0   &  8.51   &    1.97   &   0.615 & \\
353.5-04.9 & H1-36      &  5.08  &  3.50  &  7.00  &  0.03  &  0.061  & 32.0    &  0   &  0      &    0.97   &   0.659 & \\
355.1-02.9 & H1-31      &  5.04  &  3.70  &  12.0  &  0.04  &  0.11   & 21.0    &  0   &  8.77   &    1.82   &   0.633 & \\
355.1-06.9 & M3-21      &  4.95  &  3.20  &  5.00  &  0.06  &  0.11   & 18.0    &  0   &  8.65   &    3.25   &   0.613 & \\
355.4-02.4 & M3-14      &  4.90  &  3.80  &  6.00  &  0.08  &  0.35   & 23.0    &  0   &  8.79   &    3.48   &   0.609 & \\
355.7-03.5 & H1-35      &  4.65  &  4.00  &  8.00  &  0.04  &  0.14   & 10.0    &  0   &  8.27   &    3.70   &   0.593 & \\
355.9-04.2 & M1-30      &  4.60  &  3.8   &  8.00  &  0.07  &  0.22   & 22.0    &  0   &  8.76   &    3.21   &   0.594 &  wels \\
356.2-04.4 & Cn2-1      &  4.90  &  3.60  &  6.00  &  0.04  &  0.087  & 25.0    &  0   &  9.20   &    1.31   &   0.636 & \\
356.9+04.5 & M2-11      &  5.15  &  2.70  &  7.00  &  0.05  &  0.06   & 20.0    &  0   &  8.75   &    2.45   &   0.628 & \\
357.1+03.6 & M3-7       &  4.85  &  3.0   &  4.0   &  0.05  &  0.074  & 19      &  0   &  8.56   &    2.70   &   0.613 & \\
357.2+07.4 & M4-3       &  4.80  &  3.62  &  8.00  &  0.033 &  0.075  & 20.0    &  0   &  8.79   &    1.53   &   0.624 & \\
358.2+03.6 & M3-10      &  4.97  &  3.00  &  5.00  &  0.04  &  0.048  & 27.0    &  0   &  8.81   &    1.52   &   0.635 & \\
358.5-04.2 & H1-46      &  4.70  &  3.90  &  7.00  &  0.02  &  0.045  & 17.0    &  0   &  0      &    1.22   &   0.625 & \\
358.7-05.2 & M3-40      &  5.00  &  3.30  &  7.60  &  0.03  &  0.05   & 26.0    &  0   &  8.80   &    1.16   &   0.647 & \\
359.1-01.7 & M1-29      &  5.08  &  3.5   &  3.0   &  0.05  &  0.128  & 24.0    &  0   &  0      &    2.17   &   0.629 &  bipolar \\
359.2-33.5 & CRBB1      &  4.43  &  3.8   &  3.5   &  0.08  &  0.18   & 13.0    &  0   &  0      &    6.41   &   0.562 &  O(H) \\
359.3-00.9 & Hb5        &  5.08  &  3.50  &  1.30  &  0.02  &  0.03   & 25.0    &  0   &  8.84   &    0.77   &   0.668 & \\
359.7-02.6 & H1-40      &  4.85  &  3.60  &  7.80  &  0.02  &  0.04   & 21.0    &  0   &  8.45   &    1.01   &   0.642 & \\
359.8-07.2 & M2-32      &  4.90  &  3.60  &  7.00  &  0.06  &  0.19   & 29.0    &  0   &  0      &    2.15   &   0.620 & \\
           & SMP5       &  5.14  &  3.61  &  60.3  &  0.15  &  0.61   & 30.0    &  0   &  8.28   &    6.52   &   0.604 & \\
359.9-04.5 & M2-27      &  4.78  &  3.70  &  5.50  &  0.03  &  0.069  & 26.0    &  0   &  8.93   &    1.10   &   0.634 & \\
\noalign{\smallskip}
\cline{1-13}        
\end{tabular}       
\end{flushleft}     
\label{listapp2}
\end{table*}


\begin{thebibliography}{}

\bibitem{AK90}
Aaquist O.B., Kwok S., 1990, A\&AS 84, 229

\bibitem{AOS92} Acker A., Ochsenbein F., Stenholm B., et al., 1992,
Strasbourg -- ESO Catalogue of Galactic Planetary Nebulae.  ESO,
Garching (CGPN)

\bibitem{AGGD02}
Acker A., Gesicki K., Grosdidier Y., Durand S., 2002, A\&A 384, 620

\bibitem{BZ94}
Bedding T.R., Zijlstra A.A., 1994, A\&A 283, 955

\bibitem{Blo95}
Bl\"ocker T., 1995, A\&A 299, 755

\bibitem{GPGG83}
Gathier R., Pottasch S. R., Goss W. M., van Gorkom J. H.,
1983, A\&A 128, 325

\bibitem{GP89}
Gathier R., Pottasch S. R., 1989, A\&A 209, 369

\bibitem{GA96}
Gesicki K., Acker A., 1996, Ap\&SS 238, 101

\bibitem{GAS96}
Gesicki K., Acker A., Szczerba R., 1996, A\&A 309, 907

\bibitem{GZAS98}
Gesicki K., Zijlstra A.A., Acker A., Szczerba R., 1998, A\&A 329, 
265

\bibitem{GZ00}
Gesicki K., Zijlstra A.A., 2000, A\&A 358, 1058

\bibitem{GZ02}
Gesicki K., Zijlstra A.A., 2003, MNRAS 338, 347

\bibitem{G01}
G\'orny S., 2001, ApSS 275, 67

\bibitem{GST97}
G\'orny S.K., Stasi\'nska G., Tylenda R., 1997, A\&A 318, 256

\bibitem{1994} 
Habing H.J., Tignon J., Tielens A.G.G.M., 1994, A\&A
286, 523

\bibitem{M02}
Magrini L., Corradi R. L. M., Walton N. A., et al, 2002, A\&A 386, 869

\bibitem{M00}
Mellema G., 2001, ApSS 275, 147

\bibitem{NGS99}
Napiwotzki R., Green P.J., Saffer R.A., 1999, ApJ 517, 399

\bibitem{Pau88}
Pauldrach A., Puls J., Kudritzki R.P.,, et al., 1988, A\&A 207, 123

\bibitem{PSM01}
Pe\~na M., Stasi\'nska G., Medina S., 2001, A\&A 367, 983

\bibitem{Po84}
Pottasch S.R., 1984, ``Planetary Nebulae'', D.Reidel Publishing Company,
Dordrecht, Holland

\bibitem{Richer97}
Richer H.B., Fahlman G.G., Ibata R.A., et al., 1997,
ApJ, 484, 741

\bibitem{Schonberner83}
Sch\"onberner D., 1983, ApJ, 272, 708

\bibitem{SCM92}
Schwarz H.E., Corradi R.L.M., Melnick J., 1992, A\&AS 96, 23

\bibitem{ST94}
Stasi\'nska G., Tylenda R., 1994, A\&A, 289, 225

\bibitem{VW93}
Vassiliadis E., Wood P.R., 1993, ApJ 413, 641

\bibitem{Zijlstra99}
Zijlstra A.A., 1999, in: Asymptotic Giant Branch Stars
(edited by C. Waelkens), p.555

\bibitem{ZPB89}
Zijlstra A.A., Pottasch S.R., Bignell C., 1989, A\&AS 79, 329

\bibitem{ZP89} 
Zijlstra A.A., Pottasch S.R., 1989, A\&A, 216, 245

\end{thebibliography}
\end{document}